\documentclass{aastex61}

\begin{document}

\title{MagAO observations of the binary microlens OGLE-2014-BLG-1050 prefer the higher-mass solution\footnote{This paper includes data gathered with the 6.5m Magellan Clay Telescope at Las Campanas Observatory, Chile.}}

\correspondingauthor{Subo Dong}
\email{dongsubo@pku.edu.cn}

\author{Xiaojia Xie}
\affiliation{Kavli Institute for Astronomy and Astrophysics, Peking University, Yi He Yuan Road 5, Hai Dian District, Beijing 100871, People's Republic of China}
\affiliation{Department of Astronomy, Peking University, Yi He Yuan Road 5, Hai Dian District, Beijing 100871, People's Republic of China}

\author[0000-0002-1027-0990]{Subo Dong}
\affiliation{Kavli Institute for Astronomy and Astrophysics, Peking University, Yi He Yuan Road 5, Hai Dian District, Beijing 100871, People's Republic of China}

\author{Wei Zhu}
\affiliation{Canadian Institute for Theoretical Astrophysics, University of Toronto, 60 St. George Street, Toronto, ON M5S 3H8, Canada}

\author{A. Gould}
\affiliation{Department of Astronomy, Ohio State University, 140 W. 18th Avenue, Columbus, OH 43210, USA}
\affiliation{Max-Planck-Institute for Astronomy, K\"{o}igstuhl 17, D-69117 Heidelberg, Germany}

\author{A. Udalski}
\affiliation{Astronomical Observatory, University of Warsaw, Al. Ujazdowskie 4, 00-478 Warszawa, Poland}

\author{J.-P. Beaulieu}
\affiliation{School of Natural Sciences, University of Tasmania, Private Bag 37 Hobart, Tasmania 7001 Australia}
\affiliation{Sorbonne Universit\'e, UPMC Universit\'e Paris 6 et CNRS, UMR 7095, Institut d'Astrophysique de Paris, 98 bis Bd Arago, F-75014 Paris, France}

\author{L. M. Close}
\affiliation{Steward Observatory, University of Arizona, Tucson, AZ 85721, USA}

\author{ J. R. Males}
\affiliation{Steward Observatory, University of Arizona, Tucson, AZ 85721, USA}

\author{J.-B. Marquette}
\altaffiliation{(Associated to) Sorbonne Universit\'e, UPMC Universit\'e Paris 6 et CNRS, UMR 7095, Institut d'Astrophysique de Paris, 98 bis Bd Arago, F-75014 Paris, France}
\affiliation{Laboratoire d'Astrophysique de Bordeaux, Univ. Bordeaux, CNRS, B18N, all\'ee Geoffroy Saint-Hilaire, F-33615 Pessac, France}

\author{K. M. Morzinski}
\affiliation{Steward Observatory, University of Arizona, Tucson, AZ 85721, USA}

\author[0000-0003-1435-3053]{R. W. Pogge}
\affiliation{Department of Astronomy, Ohio State University, 140 W. 18th Avenue, Columbus, OH 43210, USA}
\affiliation{Center for Cosmology and AstroParticle Physics, The Ohio State University, 191 West Woodru Ave, Columbus, OH 43210, USA}

\author{J. C. Yee}
\affiliation{Center for Astrophysics $|$ Harvard \& Smithsonian, 60 Garden Street, Cambridge, MA 02138, USA}

\begin{abstract}
We report adaptive-optics (AO) follow-up imaging of OGLE-2014-BLG-1050, which is the second binary microlensing event with space-based parallax measurements. The degeneracy in microlens parallax $\pi_{\rm E}$ led to two sets of solutions, either {a $\sim(0.9,0.35)\,M_\odot$ binary at $\sim3.5$ kpc, or a $\sim(0.2,0.07)\,M_\odot$ binary} at $\sim1.1$ kpc. {We measure the flux blended with the microlensed source by conducting Magellan AO observations, and find that the blending} is consistent with the predicted lens flux from the higher-mass solution. {From the combination of the AO flux measurement together with previous lensing constraints, it is estimated that} the lens system consists of a $1.05^{+0.08}_{-0.07}\,M_{\odot}$ primary and a $0.38^{+0.07}_{-0.06}M_{\odot}$ secondary at $3.43^{+0.19}_{-0.21}$\,kpc.
\end{abstract}

\section{Background}

{Microlensing is sensitive to stellar binaries over a relatively broad range of masses and separations, and the sensitivity peaks when the projected separation is near the Einstein radius \citep{Mao:1991}}. Usually, the physical parameters including the lens mass $M_L$, the relative lens-source parallax $\pi_{\rm rel}$ and relative lens-source proper motion $\mu_{\rm rel}$ are degenerate. \citet{Gould:1992} showed that the degeneracy can be broken by measuring microlens parallax $\pi_\mathrm{E}$ and angular Einstein ring radius $\theta_\mathrm{E}$, and the lens mass can then be determined: 
\begin{equation}
M_L= \frac{\theta_\mathrm{E}}{\kappa\pi_\mathrm{E}}, 
\end{equation}
where $\kappa=8.14\,\mathrm{mas}/M_\odot$ is a constant.
Using the {\it Spitzer} space telescope, OGLE-2014-BLG-1050 (\citealt{Zhu:2015ab}, Z2015 hereafter) was the second binary microlensing event with measured space-based microlens parallax \citep{Refsdal:1966, Gould:1994aa}, after the binary microlensing event OGLE-2005-SMC-001 \citep{Dong:2007}. Unlike OGLE-2005-SMC-001, OGLE-2014-BLG-1050 showed caustic crossing features, which allowed for measuring $\theta_{\rm E}$ via finite-source effects. However, because {\it Spitzer} did not capture the caustic entrance, there were two degenerate solutions of $\pi_{\rm E}$ and therefore two solutions of the mass and distance for the lens system inferred by Z2015.

{The lens flux values expected from the two degenerate solutions (a binary with $\sim(0.9,0.35)\,M_\odot$ stars at $\sim3.5$ kpc vs. a $\sim(0.2,0.07)\,M_\odot$ binary at $\sim1.1$ kpc) are different}, and thus measuring the lens flux {allows one to distinguish one solution from the other}. Lens flux measurement also {provides one with a relation} between lens mass and distance \citep[e.g.,][]{Dong:2009aa, Beaulieu:2018aa}. In the crowded bulge field, the microlens target is usually blended with ambient stars on seeing-limited images. Diffraction-limited imaging with a space telescope such as {\it Hubble} (e.g., \citealt{Dong:2009aa}) or a large ground-based telescope with an adaptive-optics (AO) system (e.g., \citealt{Janczak:2010aa}) is needed to constrain the lens flux.

In this paper we present the observations of OGLE-2014-BLG-1050 using Magellan adaptive optics (MagAO;  \citealt{close2012,males2014,Morzinski2014}) {conducted for the lens flux measurement that leads to the resolution of the degeneracy in lensing solutions and accurate measurements of the lens parameters.}

\section{M\lowercase{ag}AO Observations}

\begin{figure}[h]
\plotone{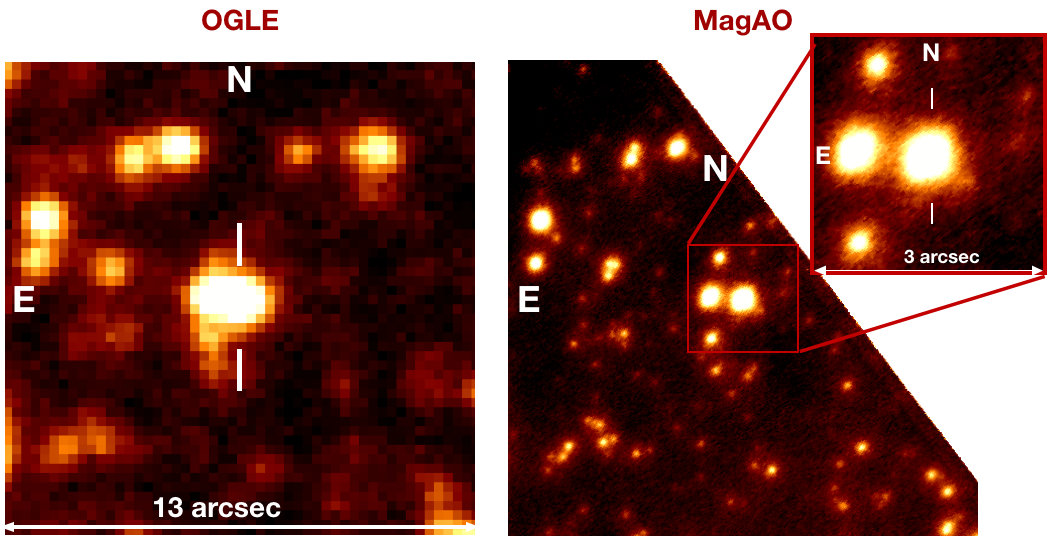}
\caption{Left: The $I$-band OGLE image with OGLE-2014-BLG-1050 at the center. Right: The $H$-band MagAO image of the same field, and the upper-right inset is a zoomed view centered on the microlens baseline object, which is indicated using a half cross. }
\label{fig:MagAO_image}
\end{figure}

We observed the field of OGLE-2014-BLG-1050 (J2000 equatorial coordinates: $\alpha=17^{\rm h}45^{\rm m}07\fs83$, $\delta=-22\degr54\arcmin20\farcs0$) using MagAO mounted on the 6.5m Magellan Clay telescope on UT 2015 May 19 in $H$ band. We used the Clio Wide camera with a pixel size of $27.5$\,mas and a field of view (FOV) of $\sim28\arcsec \times 14\arcsec$ \citep{Morzinski:2015aa}. We used an $H=10.24$\,mag AO guide star (J2000 equatorial coordinates: $\alpha=17^{\rm h}45^{\rm m}08\fs80$, $\delta=-22\degr54\arcmin33\farcs3$), which was 19\arcsec\,from the microlens. 
We took 5 frames at each of the 10 dithering positions, and the integration time for each science exposure was 30 seconds. The processing of raw images includes correcting for detector nonlinearity \citep{Morzinski:2015aa}, dark currents and flat fielding. {Then we align the frames using the {\it geomap} and {\it geotran} tasks of IRAF, which utilizes the positions of isolated bright stars as references for the alignment.} The detailed processing procedures follows Xie et al. in prep. The right panel of Figure~\ref{fig:MagAO_image} shows our processed MagAO image.

We use SExtractor \citep{Bertin:1996aa} for astrometry and aperture photometry on MagAO images. We perform astrometry calibration of MagAO images mainly following \citet[][]{Janczak:2010aa}. We first measure the position of the microlensed source on the subtracted OGLE images. Then we compute a linear coordinate transformation relation between OGLE and MagAO using the positions of seven bright and isolated stars on both sets of frames. The point source shown in the right panel of Figure~\ref{fig:MagAO_image} is identified as the microlens on the MagAO image, and its offset from the transformed OGLE source position is $\mathrm{(\Delta\alpha\, cos\delta, \Delta\delta)=(5\pm14, -5\pm10)\,{\rm mas}}$. The relative proper motion of source and lens is $\mathrm{\mu_{rel}\approx7\,mas/yr}$ (Z2015), and at the time of MagAO observations ($\sim 310$ days after the peak), the expected separation between source and lens is $\mathrm{\approx 6\,mas}$, which is much smaller than its FWHM of $250\,{\rm mas}$ on the MagAO image. The nearest detected neighbor star of the target on the MagAO image is about $900$\,mas away. Hence, the MagAO microlens identification is secure.

{We first estimate the internal photometric uncertainties by using the multiple MagAO dithers, and then we make external calibrations.} We set an aperture that is 1.5 FWHM to perform photometry of each MagAO dither and estimate the internal photometric errors by comparing the results of all dithers. The internal error of the microlens is 0.015 mag.. Due to the lack of common stars between MagAO and 2MASS, we calibrate the MagAO magnitudes using VISTA Variables in the Via Lactea (VVV) survey \citep{Minniti:2010aa} as a bridge. We use DoPhot \citep{Schechter:1993aa} to perform PSF photometry on the extracted VVV images and then calibrate to the 2MASS magnitude system using common isolated stars within $3^\prime$ of the target. Only stars with $H>13.6$ are used to avoid detector non-linearity for VVV.  We calculate the magnitude zero point of MagAO by cross matching 5 common isolated stars between MagAO and VVV images and find the standard error of the zero point calibration is 0.04\,mag. We also {use a second method to estimate the calibration error by performing photometry on a combined image by co-adding all MagAO dithers. For this method, we find a calibration error of 0.02\,mag.} The resulting photometric estimates from the two methods are consistent, although the associated uncertainties are different. Taking a conservative approach, we adopt the larger estimate. The final calibrated $H$-band flux of the microlens baseline object is $H_{\rm base}=15.78\pm0.05$.

OGLE-2014-BLG-1050 was observed by SMARTS ANDICAM \citep{ANDICAM}, which took $H$-band and $I$-band images simultaneously. We obtained the instrumental $H$-band source flux using the $I-H$ color from linear regression and the reported best-fit $I$-band source flux from Z2015. Then we calibrate the $H$-band source flux using VVV and find that  $H_s=17.98\pm0.02$. Subtracting the source flux from the baseline object, we derive the blend flux $H_{b}=15.93\pm0.06$.

\section{Constraints on lens parameters}
\label{Constraints}

We first estimate the expected $H$-band fluxes of the lens system from the two physical solutions in Z2015. We assume that both components of the binary lens are main-sequence stars, and estimate their luminosities using isochrones from Dartmouth Stellar Evolution Database \citep{Dotter:2008aa} by adopting uniform priors of [Fe/H] between -0.2 and 0.2 and age between 1 and 10\,Gyr. Following \cite{Bennett:2015aa}, we model extinction as a function of lens distance $D_L$ by

\begin{equation}
A_L=(1-\exp(-D_L/\tau_\mathrm{dust}))/(1-\exp(-D_S/\tau_\mathrm{dust}))A_\mathrm{S},
\end{equation}

where $\tau_\mathrm{dust}=0.1\,\mathrm{kpc}/\sin(b)$ is the scale height of the dust toward the galactic bulge. {There is a four-fold degeneracy of light-curve parameters \citep{Refsdal:1966}, depending on which side the observed source trajectory lies on the lens as seen from the Earth/Spitzer (i.e., the sign of impact parameters $u_0$). The four solutions are $(u_{0,\oplus}+,u_{0,Spitzer}+)$, $(u_{0,\oplus}+,u_{0,Spitzer}-)$, $(u_{0,\oplus}-,u_{0,Spitzer}+)$ and $(u_{0,\oplus}-,u_{0,Spitzer}-)$, which we denote as $(+,+)$, $(+,-)$, $(-,+)$, and $(-,-)$ below. In the case of OGLE-2014-BLG-1050, the degenerate solutions correspond to two different sets of physical lens parameters.}  The best-fit higher and lower-mass solutions in Z2015 predict $H_{L,(+,+)}=16.70\pm0.89$ and $H_{L,(-,+)}=18.51\pm0.49$,  respectively. {We note that $H_{L,(+,+)}$ is consistent with the observed blend brightness within $1\sigma$ error under the assumption that the blended light comes from the lens, but $H_{L,(-,+)}$ is inconsistent with the observed blend brightness.}

\begin{figure}[h]
\gridline{\fig{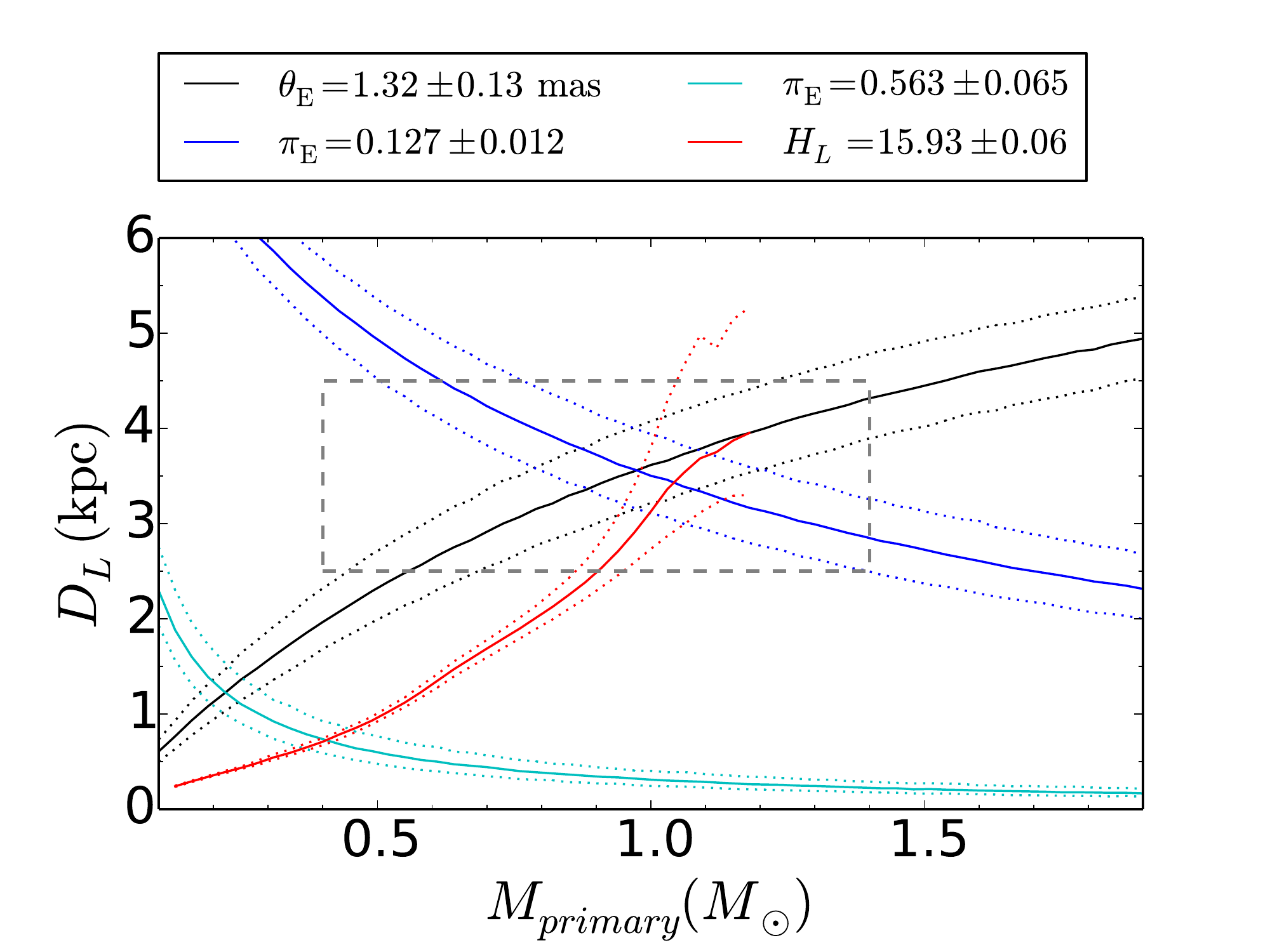}{0.55\textwidth}{}
          \fig{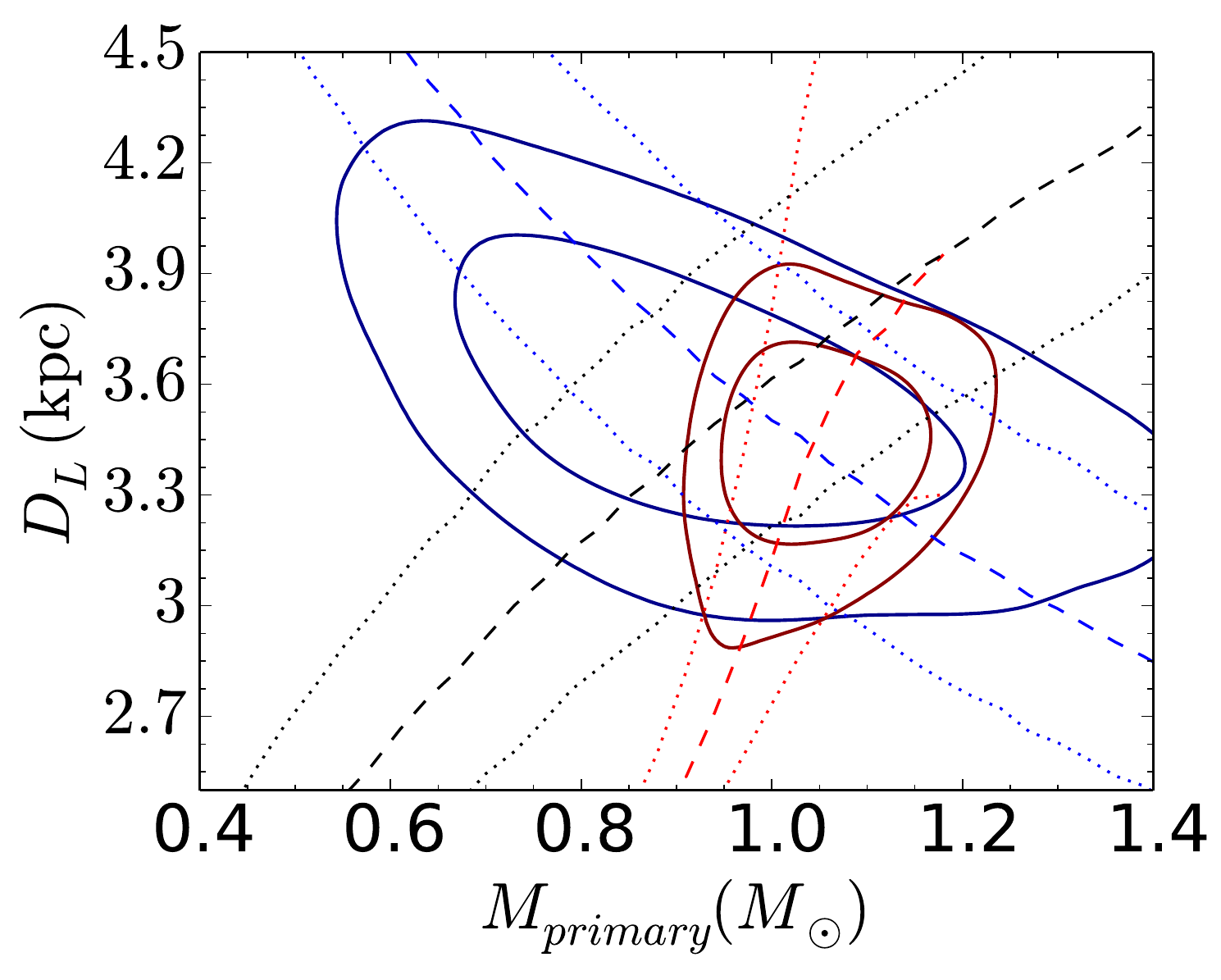}{0.45\textwidth}{}
          }
\caption{The left panel shows the constraints on lens primary mass and distance. Blue and cyan lines show the constraint from $\pi_\mathrm{E}$ of $(+,+)$ and $(-,+)$ solutions respectively. Black lines show the constraint from $\theta_\mathrm{E}$ and red line shows the constraint from lens $H$ band flux. The dotted lines show $1\sigma$ error of each constraint. The right panel is a zoom-in of the black dashed rectangle in left panel, with each constraint shown in dashed lines. The contours enclose the region with likelihood $68\%$ and $95\%$. The blue contours show the likelihood considering constraints of $\theta_\mathrm{E}$ and $\pi_\mathrm{E}$, while the dark red contours show the likelihood combining all of the three constraints $\theta_\mathrm{E}$, $\pi_\mathrm{E}$ in $(u_0+,u_0+)$ solution and lens flux.}
\label{fig:demonstration}
\end{figure}

{In Figure~\ref{fig:demonstration}, we present the constraints on the lens mass and distance given by the measured observables, including $\pi_\mathrm{E}$, $\theta_\mathrm{E}$, and $H_L$. It is found that the intersections between $\pi_\mathrm{E}$ and $\theta_\mathrm{E}$ constraints is consistent with the intersection between $\pi_\mathrm{E}$ and $H_L$ constraints for the higher-mass solution. In contrast, the $\pi_\mathrm{E}-\theta_\mathrm{E}$ and $\pi_\mathrm{E}-H_L$ intersections for the lower-mass solution shows considerable difference in the $M_\mathrm{primary}-D_L$ parameter space.} The right panel of Figure~\ref{fig:demonstration} shows the zoomed-in view in the higher-mass region, and the likelihood contours for the higher-mass solution with and without the lens flux constraint are shown in blue and dark red, respectively.

\begin{figure}[h]
\gridline{\fig{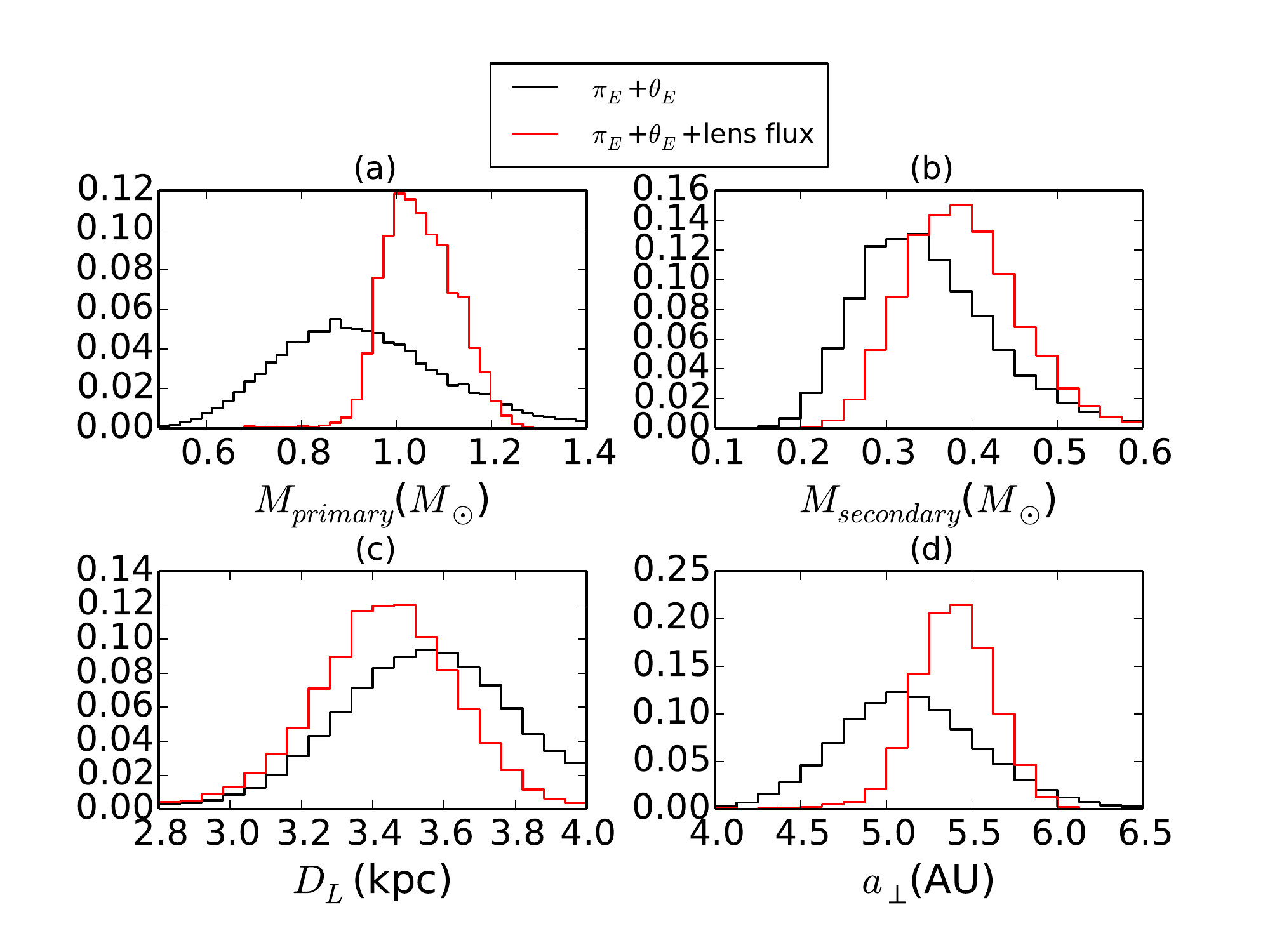}{0.9\textwidth}{}}
\caption{Red lines show the posterior distribution of (a) primary mass, (b) secondary mass, (c) lens distance and (d) projected binary separation respectively. The black lines are distributions without constraint of lens flux for comparison.}
\label{fig:histogram}
\end{figure}

Combining all the available constraints, including $\theta_\mathrm{E}$, $\pi_\mathrm{E}$ and lens flux, we estimate the physical properties of the lens system. {We note that there exist two sets of higher-mass solutions, in which both source trajectories seen from Earth and the satellite pass on the upper, $(+,+)$ solution, and lower, $(-,-)$ solution, sides with respect to the lens. We, therefore, estimate the lens parameters by weighting the relative likelihood $\exp({-\Delta{\chi^2}/2})$, where $\Delta{\chi^2}$ is the $\chi^2$ difference between the two solutions.} The posterior distributions of  physical parameters are shown in Figure \ref{fig:histogram}, showing that the binary lens system consisting of a $M_{\rm primary}=1.05^{+0.08}_{-0.07}\,M_{\odot}$ primary star orbited by a $M_{\rm secondary}=0.38^{+0.07}_{-0.06}M_{\odot}$ secondary star at a projected separation $\mathrm{5.40^{+0.23}_{-0.22}\,AU}$. The binary system is at a distance of $3.43^{+0.19}_{-0.21}$ kpc. 

\section{Discussion}

There is no other star detected between baseline object and its nearest neighbor on MagAO frame, and both stars were detected on OGLE images according to Z2015. Z2015 suggested that the OGLE blend at $I_b=17.95$ is the lens, which is expected to $I_L\sim17.9$ from the higher-mass solution, and our data provide additional support to this interpretation.

The most straightforward interpretation is that {the most of the blended flux comes from the lens.} {However, we note that it is difficult to completely} rule out (some) contributions from possible additional lens companions, source companions, or ambient stars. Given the lens-source proper motion of $\mathrm{\sim7\,mas/yr}$, the lens can be separately resolved from the source using AO on 6-10m telescopes in several years, like in the cases of \cite{Batista:2015aa} and \cite{Bennett:2019aa}. Such observations will further constrain the lens system. Moreover, because the primary is bright, further spectroscopic observations will {better constrain the lens parameters \citep[e.g.,][]{Han19}.} The expected radial velocity (RV) signals are on the order of $\sim 10\,{\rm km\,s^{-1}}$ with a period on the order of a decade, and future RV monitoring can definitively verify the lens solution and constrain its orbital properties \citep{Skowron2011, Boisse:2015aa, Santerne:2016aa, Yee:2016aa}.

\section{Acknowledgement}

X. X. and S. D. acknowledge the support of National Key
R\&D Program of China No. 2019YFA0405100 and Project 11573003 supported by NSFC. We acknowledge the Telescope Access Program (TAP) funded by NAOC, CAS, and the Special Fund for Astronomy from the Ministry of Finance.  The OGLE project has received funding from the National Science Centre, Poland, grant MAESTRO 2014/14/A/ST9/00121 to AU. KMM's work is supported by the NASA Exoplanets Research Program (XRP) by cooperative agreement NNX16AD44G.  JPB is supported by the University of Tasmania through the UTAS Foundation and the endowed Warren Chair in Astronomy. JPB and JBM acknowledge the financial support of the ANR COLD WORLDS (ANR-18-CE31-0002). 

\clearpage
\phantomsection
\bibliography{OB141050_note}

\end{document}